\documentclass[aps,pre,onecolumn,epsf,graphicx,a4]{revtex4}
\usepackage[dvips]{graphicx}
\usepackage{amsfonts}
\usepackage{epsfig}
\usepackage{amsmath}
\usepackage{amssymb}

\def\pd2x{{\partial^2 \over \partial x^2}}

\usepackage{epsfig,latexsym}

\newcommand \bew {\begin{widetext}}
\newcommand \enw {\end{widetext}}

\begin{document}

\title{\bf\noindent The statistical mechanics of multi-index matching problems 
with site disorder} 

\author{David S. Dean$^{(1)}$ and  David Lancaster$^{(2)}$ 
}

\affiliation{
(1) Laboratoire de Physique Th\'eorique, CNRS UMR 5152, IRSAMC, Universit\'e 
Paul Sabatier, 118 route de Narbonne, 31062 Toulouse Cedex 04, France\\
(2) Harrow School of Computer Science, University of Westminster, 
Harrow, HA1 3TP, UK \\
}
\date{2 July 2006}
\begin{abstract}
We study the statistical mechanics of multi-index matching problems
where the quenched disorder is a geometric site disorder rather than a
link disorder. A recently developed functional formalism is exploited
which yields exact results in the finite temperature thermodynamic
limit. Particular attention is paid to the zero temperature limit of
maximal matching problems where the method allows us to obtain the
average value of the optimal match and also sheds light on the
algorithmic heuristics leading to that optimal match.
\end{abstract}

\maketitle
\vspace{.2cm} \pagenumbering{arabic}


\section{Introduction}

Matching problems have been studied both from the perspectives of
combinatorial optimization \cite{optbook} and statistical physics
\cite{mepavi}.  In the statistical physics approach, interest lies in
computing average properties of the matching path in stochastic
versions of the problem. The most general form of the $m$-partite
multi-index problem is stated as follows. One has $m$ sets of $N$
objects and a matching consists of combining one element from each set into
a molecule or match with $m$ components. If in the $i$-th match 
the element we choose from set  $j$, $1\leq j\leq m$, is denoted by $i_j$, 
then the cost of the $i$-th match is given by the entry $C_{i_1\cdots i_m}$
of the cost matrix. The total cost of the matching is then
given by the sum of the costs of the $N$ individual matches. The 
underlying maximum/minimum optimization problem consists of finding the 
global match which maximizes/minimizes this cost.
The bipartite matching problem with $m=2$ has been most studied
and some analytic
progress has been made for so called ``random link'' models
where the elements of the cost matrix
$C$ are chosen independently from some probability distribution
 \cite{mepavi,ralink,mameri}.

Often matching problems have a geometrical origin as the points to be
matched lie in some $d$-dimensional domain $\cal{D}$. Euclidean
matching problems are of practical importance, finding applications in
air traffic control \cite{atraf} and the manufacture of printed
circuits \cite{manu}. The three dimensional Euclidean matching
problem, both minimal and maximal has been studied numerically via
Monte Carlo simulations and in particular simulated annealing
\cite{num3d}.  In these geometric problems the matching matrix $C$ is
determined from the location of the points using some metric on the
space.  As a result matrix elements are no longer independent (for
example triangle inequalities exist for the case when the matching
matrix depends on the Euclidean distance between the points). The
random link versions can be regarded as approximations to Euclidean
models, and in minimal bipartite matching this approximation seems
rather good \cite{mfeucl} and corrections to account for triangular
correlations have been computed \cite{ralink}.

Recently, we have developed new techniques for analyzing Euclidean
traveling salesman type problems (TSP) \cite{us} based on functional
integration and a functional order parameter. This formalism does not
yield the average value of the optimal path for the minimal TSP
because it does not apply when the energy is rescaled to make the
ground state energy extensive. However in the case of maximal TSP type
problems the ground state energy is extensive without any energy
rescaling and the average value of the optimal solution can be
computed exactly. The method \cite{us} does not rely on the replica
trick, however the results obtained show that for the Euclidean TSP
the system is replica symmetric, as is the case for random link
calculations \cite{mepavi}. In this paper we show how the the
formalism of \cite{us} can be applied to multi-index matching problems
at finite temperature and also how it can be used to obtain the
average value of the optimal match in the maximal case. In addition
the method gives us information concerning the algorithmic heuristic
leading to the optimal path. The mathematical structure of the
multi-index matching problem in our thermodynamic formalism is more
complicated than that of the Euclidean TSP in that a functional order
parameter appears for each of the $m$ sets of points. However in the
cases studied here the functional order parameters are the same for
each set and we find no evidence of symmetry breaking. As a
consequence of this symmetry we find that the thermodynamics of the
bipartite matching problem and the Euclidean TSP are essentially
equivalent. For certain other multi-index problems, with chain-like
matching functions, we find relations between the average path lengths
computed in these models and the TSP.

The statistical mechanical cavity approach has recently been applied to
minimal multi-index matching problems in the context of independent
links \cite{mameri}. In this problem the energy (or equivalently the
temperature) is scaled to ensure that the ground state energy,
corresponding to the maximal match, is extensive.  In contrast to
the bipartite matching problem which has a replica symmetric solution,
it was shown that the multi-index matching problem for more than two
indices has a low temperature glassy phase characterized by replica
symmetry breaking. It is therefore interesting to see if the formalism 
developed in \cite{us} which is exact for maximal optimization problems 
shows similar behavior when applied to multi-index matching. However in
all of the cases studied here the system exhibits no such phase transition,
suggesting that they appear to be non-glassy  from the statistical mechanics 
point of view.

\section{Model and general solution}

Here we define the general multi-index matching problem with site disorder.

\begin{figure}
\epsfxsize=0.3\hsize \epsfbox{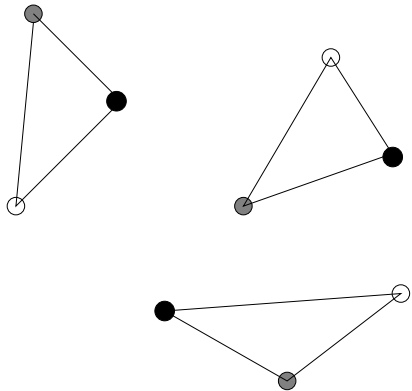}
\caption{Full match for three sets (white, black, gray) of three points 
($m=3$, $N=3$).} 
\label{figfmatch}
\end{figure}

Consider the following $m$-partite matching problem with $m= K+1$ sets
$S^{(0)}, S^{(1)}, \cdots, S^{(K)}$, each set $S^{(a)}$ consists of
$N$ points $\{{\bf r}^{(a)}_1, \cdots,{\bf r}^{(a)}_N \}$ distributed
in a common domain $\cal D$ in a space of dimension $d$. We assume
that the points are independently distributed within the domain and
that those in set $S^{(a)}$ are distributed according to a probability
density $p_a$. An individual match consists of $K+1$ points where one
point in each set is matched with one point in each other set. Each
match has an energy function associated with it which we denote by
$V({\bf r}^{(0)},{\bf r}^{(1)}, \cdots {\bf r}^{(K)})$.  The
multi-index optimization problem involves making $N$ individual
matches (Fig. (\ref{figfmatch})),
 where each point is associated with one and only one matching,
and optimizing the total match energy.  The case of $K=1$ corresponds
to the well known bipartite matching problem.  A micro-state of the
model is specified by $\{(i, \sigma^{(1)}_i \cdots \sigma^{(K)}_i)\}$
where the point $i$ of the set $S^{(0)}$ having position ${\bf
r}_i^{(0)}$ ($1\leq i\leq N)$) is matched with ${\bf
r}_{\sigma^{(1)}_i}^{(1)}\cdots, {\bf r}_{\sigma^{(K)}_i}^{(K)}$ in
the other sets and $\sigma^{(a)}_i$ denotes the label of the element
in set $a$ chosen to be matched with the element $i$ in set $0$.  The
$\sigma^{(a)}$ are permutations on the $N$ elements of $S^{(a)}$. The
phase space for multi-index matching is thus the Cartesian product
$(\Sigma_N)^K$, where $\Sigma_N$ is the permutation group on $N$
elements. The size of this phase space is $(N!)^K$ and the entropy is
consequently super-extensive.

The Hamiltonian of the system is given by adding up the energy for each 
individual match
\begin{equation}
H[\sigma_1,\sigma_2,\cdots,\sigma_K] = 
\sum_{i=1}^N V({\bf r}_i^{(0)},{\bf r}_{\sigma^{(1)}_i}^{(1)},\cdots,
 {\bf r}_{\sigma^{(K)}_i}^{(K)} ),
\end{equation}
and the partition function for sets of $N$ points is given by
\begin{equation}
Z_N =\sum_{\sigma^{(1)},\sigma^{(2)},\cdots \sigma^{(K)}}
\exp\left(-\beta H[\sigma_1,\sigma_2,\cdots\sigma_K]\right),
\end{equation}
where $\beta = 1/T$ and  $T$ is the canonical temperature of the system.

\begin{figure}
\epsfxsize=0.2\hsize \epsfbox{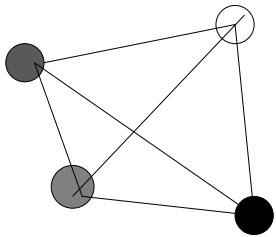}
\caption{Individual match between four points (each in a different set white,
gray, black and stripes), showing a fully symmetric and fully connected cost
function.} 
\label{figtcchain}
\end{figure}

The form of $V$ can be arbitrary, but a symmetric potential function
where $V({\bf r}_0,\cdots,{\bf r}_K)$ is left invariant by any
rearrangement of its arguments is natural in this context.  A simple
way of constructing such a potential is to consider the potential to
be the, {\em totally connected}, sum of pairwise matching potentials between all pairs as shown
in Fig. (\ref{figtcchain}):
\begin{equation}
V({\bf r}_0,{\bf r}_1\cdots{\bf r}_K) = 
\sum_{a=1,b=1}^K V_P({\bf r}_{a},{\bf r}_b).
\label{pairwise}
\end{equation}

\begin{figure}
\epsfxsize=0.2\hsize \epsfbox{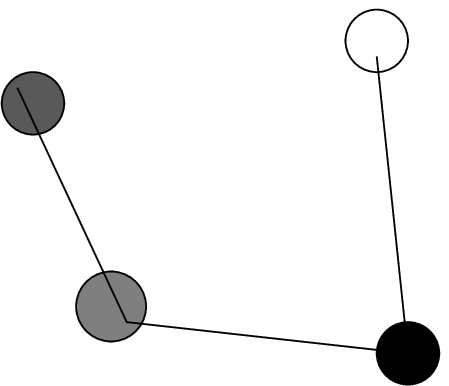}
\caption{Individual match between four points (each in a different set white,
gray, black and stripes), showing a open chain cost function.} 
\label{figopenchain}
\end{figure}

\begin{figure}
\epsfxsize=0.2\hsize \epsfbox{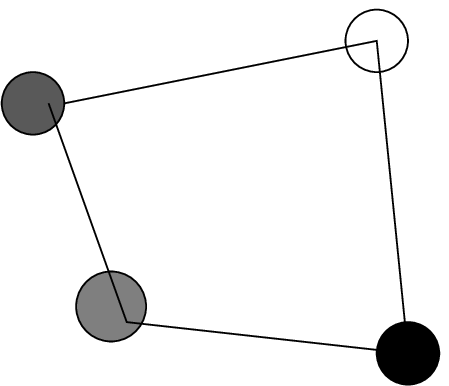}
\caption{Individual match between four points (each in a different set white,
gray, black and stripes), showing a closed chain cost function.} 
\label{figclosedchain}
\end{figure}

However, to make analytic progress, much of our analysis
considers potentials that are pairwise but are not symmetric.
These potentials allow the match to be represented as an ordered path through
points from each successive set
as  illustrated in  Fig. (\ref{figopenchain}).
\begin{equation}
V({\bf r}_0,{\bf r}_1\cdots{\bf r}_K) = 
\sum_{a=1}^K V_P({\bf r}_{a-1},{\bf r}_a).
\label{openpairwise}
\end{equation}
Besides this {\em open} model, a generalization is to add 
$V_P({\bf r}_{K},{\bf r}_0)$, to arrive at the situation shown 
in Fig. (\ref{figclosedchain}) and we shall term this
a {\em closed} or {\em cyclic} model.

Because we emphasize the geometric interpretation of
pairwise matching,
we will frequently take the pair potential $V_P$ to 
be the Euclidean distance $V_P({\bf r}_{i-1},{\bf r}_i)
=|{\bf r}_{i-1}-{\bf r}_i|$, though we also treat the quadratic form 
which seems more amenable to analytic approaches. 
The case $K=1$ corresponds to the Euclidean bipartite matching problem.
The cyclic, $K=2$ tripartite case is symmetric and corresponds to the 
{\em triangle} interaction which amounts to
the total length needed to form a triangle between three points. 
Another symmetric possibility for $K=2$, tripartite matching, that is not
of the pairwise form, is the {\em star} interaction 
where the three points are 
connected by three lines emanating from the point which would be their
center of mass (for equal masses)
\begin{equation}
V_{star}({\bf r}_0, {\bf r}_1, {\bf r}_2) = |{\bf r}_0-{\bf R}| +
|{\bf r}_1-{\bf R}| + |{\bf r}_2-{\bf R}|,
\end{equation}
where 
\begin{equation}
{\bf R} = {1\over 3}({\bf r}_0 + {\bf r}_1+{\bf r}_2 ).
\end{equation}


As pointed out in \cite{us} the disorder in this type of 
problem can be encoded
in the unaveraged density of the quenched points in the domain ${\cal D}$
\begin{equation}
\rho_a({\bf r}) = {1\over N}\sum_{i=1}^N 
\delta\left({\bf r} -{\bf r}_i^{(a)}\right),
\end{equation}
note that the value  of $\rho_a$ averaged over the disorder is 
by definition $p_a$.

Imposing this constraint we may discard the sum over permutations and write
\begin{equation}
Z_N = C_N \int \prod_{i,a} d{\bf r}_i^{(a)}
{\prod_{\bf r}}\prod_a \delta\left[N \rho_a({\bf r})- 
\sum_{i=1}^N \delta({\bf r}_i^{(a)} - {\bf r})\right]
\exp\left(-\beta \sum_{i} V({\bf r}_i^{(0)}, \cdots,{\bf r}_i^{(K)})\right).
\end{equation}
In the above formulation without the delta function constraints the 
matches chosen can use any point in the domain $\cal D$, however the delta
function constraint allows the use of only the points which are available
in the sets $S_a$. The delta function constraints which are present at each 
point ${\bf r}$ in the domain $\cal{D}$ are now expressed as a 
functional Fourier integral (the $\mu$ integration is along the imaginary axis)
\begin{equation}
Z_N = C'_N \int \prod_{a} d[\mu_a]\prod_{i,a} d{\bf r}_i^{(a)}
\exp\left(N\int d{\bf r} \sum_a \mu_a({\bf r}) \rho_{a}({\bf r})\right)
\exp\left(-\sum_{a i} \mu_a({\bf r}_i^{(a)})-\beta \sum_{i} V({\bf r}_i^{(0)}, \cdots,{\bf r}_i^{(K)})\right).
\end{equation}
The integrals over the dynamical variables ${\bf r}_i^{(a)}$ which are
taken to be in $\cal D$ may now be carried out independently ,
changing the normalization and leaving an integration over only $K+1$
functional variables
\begin{equation}
Z_N = C''_N\int  \prod_{a} d[\mu_a]
\exp\left[N\left( \int d{\bf r} \sum_a \mu_a({\bf r})\rho_{a}({\bf r})
+ \ln\left(\int \prod_a d{\bf r}^{(a)}\exp\left(-\sum_a \mu_a({\bf r}^{(a)})
-\beta V({\bf r}^{(0)}, \cdots {\bf r}^{(K)})\right)\right)\right)\right].
\end{equation}
The terms $C_N,C_N',C_N''$ are all constants whose value is unimportant.
Now in the limit of large $N$ we use that fact that
\begin{equation}
\int d{\bf r} \mu_a({\bf r}) \rho_a({\bf r}) \to
 \int d{\bf r} \mu_a({\bf r}) p_a({\bf r}) + O({1\over \sqrt{N}}),
\end{equation}
to eliminate the dependence on the quenched disorder in $\rho_a({\bf r})$,
and obtain

\begin{equation}
Z_N \approx C''_N\int  \prod_{a} d[\mu_a]
\exp\left[N\left( \int d{\bf r} \sum_a \mu_a({\bf r})p_{a}({\bf r})
+ \ln\left(\int d{\bf r}^{(a)}\exp\left(-\sum_a \mu_a({\bf r}^{(a)})
-\beta V({\bf r}^{(0)}, \cdots {\bf r}^{(K)})\right)\right)\right)\right].
\end{equation}

The above integral may be evaluated by the saddle point method, the method 
works in this limit because the thermodynamics is determined solely by
the $p_a$ which are the first moments of the random quenched  densities 
$\rho_a$. We find the following expression for the free energy per number of 
particles, $N$, in each set $S_a$

\begin{equation}
-\beta f = 
{\rm max}_{\mu_a}  \left[ \int d{\bf r} \sum_a \mu_a({\bf r})p_{a}({\bf r})
+ \ln\left(\int d{\bf r}^{(a)}\exp\left(-\sum_a \mu_a({\bf r}^{(a)})
-\beta V({\bf r}^{(0)}, \cdots {\bf r}^{(K)})\right)\right)\right] + {\gamma_N
\over N}.
\end{equation}
The last term is a beta and $\mu$ independent term and 
is related to the infinite 
temperature entropy, it does not contribute to the energy of the 
system, which is what interests us here.
The resulting saddle point equation for the $\mu_a$ is

\begin{equation}
p_a({\bf r}_a) = {1\over \cal Z}\exp(-\mu_a({\bf r}_a))\int \prod_{b\neq a} 
d{\bf r}^{(b)}\exp\left(-\sum_{b\neq a} \mu_b({\bf r}^{(b)})
-\beta V({\bf r}^{(0)}, \cdots,{\bf r}^{(a)}\cdots, {\bf r}^{(K)})\right),
\end{equation}

where 

\begin{equation}
{\cal Z} = \int d{\bf r}^{(a)}\exp\left(-\sum_a \mu_a({\bf r}^{(a)})
-\beta V({\bf r}^{(0)}, \cdots {\bf r}^{(K)})\right).\label{eqcalz}
\end{equation}

One sees in the above that if $\mu_a({\bf r})$ is a solution to the saddle
point equation then so is $\mu_a({\bf r}) + \alpha_a$, where 
$\alpha_a$ are arbitrary
constants. This is because the constraints $\int d{\bf r} \;p_a({\bf r})
= 1$ are automatically satisfied and so the zero frequency Fourier modes of 
the $\mu_a$ are redundant and are zero modes of the theory. We can use one
of these free zero modes to set $\cal Z$ in  Eq. (\ref{eqcalz}) to be equal 
to one. The saddle point equation then 
becomes

\begin{equation}
p_a({\bf r}^{(a)}) = \exp\left(-\mu_a({\bf r}^{(a)})\right)\int \prod_{b\neq a} 
d{\bf r}^{(b)}\exp\left(-\sum_{b\neq a} \mu_b({\bf r}^{(b)})
-\beta V({\bf r}^{(0)}, \cdots,{\bf r}^{(a)},\cdots, {\bf r}^{(K)})\right),
\label{eqsp}
\end{equation}
and we find that the part of the free energy relevant to the calculation 
of the energy is
\begin{equation}
\beta f = -\int d{\bf r} \sum_a p_a({\bf r}) \mu_a({\bf r}), \label{eqfe}
\end{equation}
and the average energy is given by $\epsilon = 
\partial \beta f/\partial \beta$.  
This can be manipulated to find that in general, the
energy may be written:
\begin{equation}
\epsilon = \int \prod_{a} d{\bf r}^{(a)} 
V({\bf r}^{(0)}, \cdots,{\bf r}^{(a)},\cdots, {\bf r}^{(K)})
\exp\left(-\sum_{b} \mu_b({\bf r}^{(b)})
-\beta V({\bf r}^{(0)}, \cdots,{\bf r}^{(a)},\cdots, {\bf r}^{(K)})\right).
\end{equation}

We have verified the formalism
for non-uniform distributions in some simple cases, but the rest
of the paper will be concerned with  uniform 
distributions in domains of unit size. 
We  write $s_a({\bf r})= \exp({\mu_a({\bf
r})})$ to obtain the equations in the following form
\begin{eqnarray}
s_a({\bf r}^{(a)}) &=& \int \prod_{b\neq a} 
d{\bf r}^{(b)}
\exp\left(-\beta V({\bf r}^{(0)}, \cdots,{\bf r}^{(a)},\cdots, {\bf r}^{(K)})\right)
\prod_{b\neq a} {1\over s_b({\bf r}^{(b)})} \nonumber\\
\epsilon &=& \int \prod_{a} d{\bf r}^{(a)} 
V({\bf r}^{(0)}, \cdots,{\bf r}^{(K)})
\prod_{b\neq a} {1\over s_b({\bf r}^{(b)})} 
\exp\left(
-\beta V({\bf r}^{(0)},\cdots,{\bf r}^{(K)})\right).
\label{eqsess}
\end{eqnarray}
These equations now form the basis for the rest of the paper
that explores their consequences. 

In general Eq. ({\ref{eqsp}) cannot be solved analytically, however in the 
limit $\beta \to \infty$ it can be simplified by a saddle point method
if one writes $s_a({\bf r}) = \exp(-\beta w_a({\bf r}))$ which yields
\begin{equation}
w_a({\bf r}) = {\rm min}_{{\bf r}_1\cdots {\bf r}_K|  {\bf r}_a = {\bf r}}
\{ V({\bf r}_1,\cdots, {\bf r}_a) - \sum_{b\neq a} w_b({\bf r}_b).
\} 
\label{eqzerotemp}
\end{equation}
When we are interested in the  maximum rather than the  minimal problem
we can either change the sign of $V$ or $\beta$, changing the sign of $\beta$
simply leads  the $\min$ above to be changed to a $\max$.
In the limit $\beta\to\infty$, using Eq. ({\ref{eqfe}), this yields the 
ground state energy or optimal matching cost to be 
\begin{equation}
\epsilon_{GS}= \sum_a\int d{\bf r} \; w_a({\bf r}).
\label{eqzerotempenergy}
\end{equation}

The saddle point equations (\ref{eqsess}) 
are similar to those studied in our work
on Hamiltonian paths but with the major difference that there are now
$K+1$ functional order parameters $\mu_a({\bf r})$. In the case where
$V$ is a symmetric function and all the $p_a$ are the same for each set,
implicitly the case here as we take them all to be uniform, there will
clearly be a {\em set symmetric solution} where $\mu_a({\bf r})= \mu({\bf r})$.
However it is possible that this symmetry could be spontaneously broken. Thus,
despite the fact that the matching problem superficially looks somewhat
simpler than  the Hamiltonian path problem, it has the potential to exhibit
more complex behavior. 

The simplest example we can consider is the bipartite matching problem
with $K=1$. In this case the saddle point equations read
\begin{eqnarray}
s_0({\bf r})
& =& \int d{\bf r}' {\exp\left(
-\beta V({\bf r},{\bf r}')\right)\over s_1({\bf r}')}\nonumber \\
s_1({\bf r})
& =& \int d{\bf r}' {\exp\left(
-\beta V({\bf r}',{\bf r})\right)\over s_0({\bf r}')}.
\end{eqnarray}
If one considers the set symmetric solution $s_0({\bf r})= s_1({\bf r})= 
s_{tsp}({\bf r})$, the resulting equation is 
\begin{equation}
s_{tsp}({\bf r})
 = \int d{\bf r}' {\exp\left(
-\beta V({\bf r},{\bf r}')\right)\over s_{tsp}({\bf r}')},
\end{equation}
this is exactly the same equation as that occurring for the TSP problem
however in the TSP problem 
$s_{tsp}({\bf r}) =\exp\left(\mu({\bf r})/2\right)$, where
$\mu$ is again the Lagrange multiplier fixing the density of points and 
the TSP free energy per site  is given by
\begin{equation}
\beta f_{tsp} = -\int d{\bf r} \mu({\bf r}) = -2 \int d{\bf r} \;
\ln\left(s_{tsp}({\bf r})\right) .
\end{equation}
However we see immediately that for the bipartite matching problem the 
free energy, within the set symmetric solution is given by
\begin{equation}
\beta f =  -2 \int d{\bf r} \;
\ln\left(s({\bf r})\right)  = \beta f_{tsp}.
\end{equation}
Thus the free energy of the TSP with $N$ points is the same as that of a 
bipartite match between two sets of $N$ points. Notice that in the 
TSP there are $N$ links and in the bipartite match there are also 
$N$ links, this equivalence is conjectured to hold  for the 
two minimal versions in the limit where the energy is scaled so
as to become extensive as $\beta\to \infty$. If the 
set symmetric solution is indeed that valid everywhere, our result
shows the strict thermodynamic equivalence of the two models at 
any finite temperature and for an arbitrary distance function !

For pairwise potentials arranged as paths Eq. (\ref{openpairwise}) and uniform 
probability density over the domain we may search for 
general $K$ solutions along the lines of the one found above for the
bipartite case. The non-cyclic case turns out to be more tractable.
We search for a solution in which 
$s_a({\bf r})= \exp({\mu_a({\bf r})})$ is the same for all
sets except the first ($a=0$) and last ($a=K$) where the
ends of the matching path lie.
\begin{equation}
s^{(a)}({\bf r}) = 
\left\{
\begin{array}{ll}
s({\bf r}) &\mbox{$a=1\dots K-1$}\\
s_E({\bf r}) &\mbox{$a=0,K$}
\end{array} 
\right.
\end{equation}
The saddle point equations now read
\begin{eqnarray}
s_E({\bf r}_0)
& =& \int d{\bf r}_1\dots d{\bf r}_K
 {\exp\left(
-\beta \sum_{a=1}^K V_P({\bf r}_{a-1},{\bf r}_a)
\right)
\over s({\bf r}_1)\dots s({\bf r}_{K-1})s_E({\bf r}_{K})
}\\
s({\bf r}_1)
& =& \int d{\bf r}_0 d{\bf r}_2 \dots d{\bf r}_K
 {\exp\left(
-\beta \sum_{a=1}^K V_P({\bf r}_{a-1},{\bf r}_a)
\right)
\over s_E({\bf r}_0) s({\bf r}_{2})\dots s_E({\bf r}_{K})
},
\end{eqnarray}
and free energy corresponding to this saddle point is 
\begin{eqnarray}
\beta f &=& -\int d{\bf r} \;\left[2\ln( s_E) + (K-1)\ln(s)\right].
\end{eqnarray}
Using the structure of the pairwise potential,
these equations can be solved in terms of a set of $K$
coupled integral equations for quantities $t_a({\bf r})$
\begin{eqnarray}
t_0({\bf r})
& = & \int d{\bf r}' 
 \exp\left(
-\beta V_P({\bf r},{\bf r}')\right)
{1\over t_{K-1}({\bf r}')}
= 
T({1\over t_{K-1}})\\
t_1({\bf r})
& = & \int d{\bf r}' 
 \exp\left(
-\beta V_P({\bf r},{\bf r}')\right)
{1\over t_{K-2}({\bf r}')}
= 
T({1\over t_{K-2}})\\
t_2({\bf r})
& = & \int d{\bf r}' 
 \exp\left(
-\beta V_P({\bf r},{\bf r}')\right)
{t_0({\bf r}') \over t_1({\bf r}') t_{K-2}({\bf r}')}
= 
T({t_0 \over t_1 t_{K-2}} )\\
t_3({\bf r})
& = & \int d{\bf r}' 
 \exp\left(
-\beta V_P({\bf r},{\bf r}')\right)
{t_0({\bf r}') \over t_2({\bf r}') t_{K-2}({\bf r}')})
= 
T({t_0 \over t_2 t_{K-2}})\\
t_{K-2}({\bf r})
& = & \int d{\bf r}' 
 \exp\left(
-\beta V_P({\bf r},{\bf r}')\right)
{t_0({\bf r}') \over t_{K-3}({\bf r}') t_{K-2}({\bf r}')}
= 
T({t_0\over t_{K-3} t_{K-2}} )\\
t_{K-1}({\bf r})
& = & \int d{\bf r}' 
 \exp\left(
-\beta V_P({\bf r},{\bf r}')\right)
{1\over t_0({\bf r}')}
= 
T({1\over t_0}),
\label{opents}
\end{eqnarray}
where we have introduced $T$ as the integral operator
appearing throughout. Once the $t_a$'s are determined
the solution is given by:
\begin{eqnarray}
s_E({\bf r}) &=& t_{K-1}({\bf r})\\
s({\bf r}) &=& t_0({\bf r})t_{K-2}({\bf r}).
\end{eqnarray}
This set of equations always has a solution with all $t_a$'s taken
the same $t_a({\bf r}) = s_{tsp}({\bf r})$, this then yields
\begin{eqnarray}
s_E({\bf r}) &=& s_{tsp}({\bf r})\\
s({\bf r}) &=& s_{tsp}^2({\bf r}),
\end{eqnarray}
and inserting this into the free energy we have
\begin{equation}
\beta f = -2 K \int d{\bf r} \ln(s_{tsp}) = K\beta f_{tsp}.
\end{equation}
The interpretation is clearly that the free energy per link of this
matching problem is exactly the same as for the TSP. This result is
supported by direct Monte Carlo simulation in the case of a two
dimensional box. This is a generalization of the thermodynamic
equivalence seen between the TSP and the bipartite matching problem.
Unfortunately, neither for closed paths nor for fully connected
matches have we found any general results and we shall consider only
special cases in the following section. Finally we comment that for
sufficiently symmetric potentials on closed domains (e.g. with
periodic boundary solutions) we find the solution $s_a = Constant$ for
all $a$. This means that the annealed approximation to free energy is
exact \cite{us} and it is easy to show that $\epsilon_{GS}$ is just
the ground state energy of a single molecule or match with $m=K+1$
sites in the case where each of the $m$ sites is free to move.

\begin{figure}
\epsfxsize=0.5\hsize \epsfbox{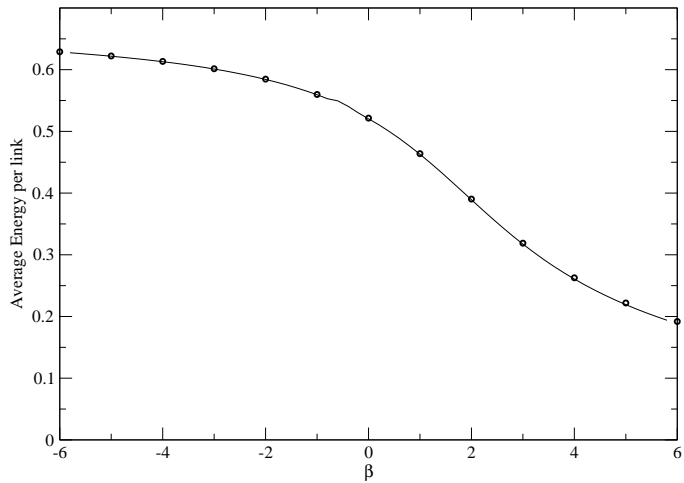}
\caption{Tripartite matching with a triangular potential
in 2 dimensions. Points are from Monte Carlo (the error bars are
too small to show) and the continuous line is an iterative solution
of Eq. (\ref{eqsp}). 
Negative temperatures correspond to the maximal problem.}
\label{fig:montecarlo}
\end{figure}

\section{Tripartite matching- numerical results}

Our analytic efforts have not revealed a solution for 
a symmetric $K=2$ tripartite system other than a constant on
symmetric closed domains.
We have therefore performed
some numerical simulations of the triangle potential.
We have considered two domains each with boundaries: a one dimension
line and in a 2D box. After preliminary tests to estimate the
strength of $1/N$ corrections and the time necessary to
equilibrate we choose the following Monte Carlo parameters:
$N=1000$, equilibration for one million steps and measurements taken over
the succeeding one million steps. Energies are averaged over 
100 samples with different sets of random points.
Our codes are checked by showing that the bipartite $K=1$ case
agrees with the TSP result. 

In Fig. (\ref{fig:montecarlo})
the Monte Carlo results for the energy are confronted
with a numerical (iterative) solution of the equations. 
We find that in all cases, the iterative approach converges
to a symmetric solution with all $s_a({\bf r})$'s the same,
even when each $s_a({\bf r})$ is seeded with rather different 
initial conditions. In two dimensions, the iterative solution
is slow for a reasonable discretization of the domain, however
we find that for the temperatures we consider the solutions are
smooth and the procedure gives good results even for a
discretization with only 400 points.

We have also considered tripartite matching with the star potential
and again see no evidence for breaking of the symmetry between
the sets.

\begin{figure}
\epsfxsize=0.4\hsize \epsfbox{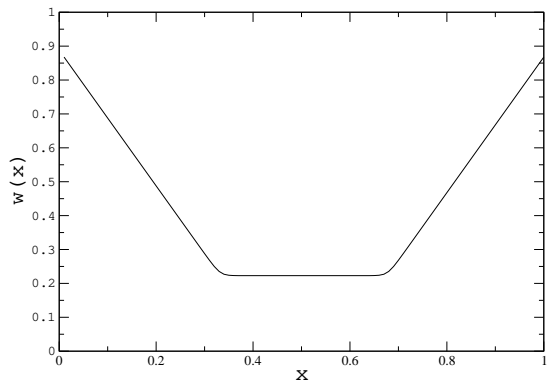}
\caption{The function $w(x)$ for tripartite matching on a one
dimensional line. Obtained by numerically solving Eq. (\ref{eqsess})
for $\beta=70.0$. The parameters agree with the solution based on 
a greedy heuristic discussed in the text.}
\label{figmaxtripartite}
\end{figure} 

\begin{figure}
\epsfxsize=0.3\hsize \epsfbox{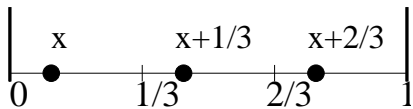}
\caption{Arrangement of points for the one third jump heuristic.}
\label{figthirdjump}
\end{figure} 

\begin{figure}
\epsfxsize=0.3\hsize \epsfbox{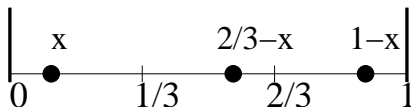}
\caption{Arrangement of points for the greedy heuristic.}
\label{figgreedy}
\end{figure} 
 
\section{Zero temperature results in one and two dimensions and 
corresponding heuristics}

In this section we consider one and two dimensional examples where our
formalism can solve the original maximal optimization problem.  We
study Eqs. (\ref{eqzerotemp}), (\ref{eqzerotempenergy}) and find that,
as was the case for the TSP \cite{us}, the method of solution throws
light on local heuristics which might be used to solve specific
instances.

First consider maximal tripartite matching in 1 dimension with a triangular potential
based on a pairwise potential that is simply the distance between
the points. 
\begin{equation}
V_{triangle}(x_0, x_1, x_2) = |x_0-x_1| +
|x_1-x_2| + |x_2-x_0|
= 2\left[ \max(x_0, x_1, x_2)-\min(x_0, x_1, x_2)\right].
\label{linm}
\end{equation}
Evidently the location of the middle point does not contribute to 
the potential. We search for a piecewise linear function $w(x)$
consisting of three linear sections evenly dividing the unit interval
as shown in Fig. (\ref{figmaxtripartite}).
A 1/3rd jump heuristic guesses that the maximum occurs for points
arranged as shown in Fig. (\ref{figthirdjump}). This basically means that
the match consists of a central point in the central third of the interval, 
matched with points in the first and last third, each  at a distance of 
a third from the central point. In this case the Eq. (\ref{eqzerotemp}) can be written:
\begin{equation}
w(x)+w(x+1/3)+w(x+2/3) = V_{triangle}(x, x+1/3, x+2/3) = 4/3, \label{cmass}
\end{equation}
where $x: 0<x<1/3$. This of course immediately gives the average value
of $4/3$ per match in the optimal match. We now make a piecewise linear anzatz 
for  $w(x)$ (which is symmetric about $x={1\over 2}$)
\begin{eqnarray}
w(x) &=& a |x-{1\over 2}| + b \ \ {\rm for} \ x\in[0,{1\over 3}]\cup  [{2\over 3},1
]\nonumber \\
 &=& a' |x-{1\over 2}| + b' \ \ {\rm for}\  x\in[{1\over 3},{2\over 3}].
\end{eqnarray}
Immediately one sees from Eq. (\ref{cmass}) that we mus have $a'=0$. 
Now substituting this into the  Eq. (\ref{eqzerotemp})  we find 
\begin{equation}
{4\over 3} a + 2b + b' = {4\over3},
\end{equation}
The coefficient  $a$ can be determined by looking at locally at Eq. (\ref{eqzerotemp}) when 
$x\in [{1\over 3}, {2\over 3}]$ and writing $x_1 = x -{1\over 3} +\epsilon_1$
and $x_2 = x +{1\over 3} +\epsilon_2$, this gives
\begin{equation}
b' = \max_{\epsilon_1,\epsilon_2} \left\{ {4\over 3} + 
2 (\epsilon_2 -\epsilon_1) -a (\epsilon_2-\epsilon_1) -{2\over 3} a - 2 b
\right\}
\end{equation}
and the local stationarity with respect to $\epsilon_1$ and $\epsilon_2$ then
gives us that $a=2$. Now we assume continuity of $w$ which gives
\begin{equation}
b' = {2\over 9}
\end{equation}
This thus yields the solution for Eq. (\ref{eqzerotemp}) in complete
agreement with the numerical resolution of the low temperature saddle
point equations shown in Fig. (\ref{figmaxtripartite}).

An alternative heuristic for this triangular distance potential is the
greedy heuristic with the arrangement of points shown in
Fig. (\ref{figgreedy}). In this case the maximization equation is:
\begin{equation}
w(x)+w(2/3-x)+w(1-x) = V_{triangle}(x, 2/3-x, 1-x) = 2(1-2x).
\end{equation}
Interestingly despite the fact that the heuristic here is different it yields
the same function $w(x)$, and consequently the same average value per match
of the optimal match. This is presumably a peculiarity of linear pairwise
potentials in one dimension and a similar phenomenon is seen for the Maximal 
TSP \cite{us}.

These issues generalize to other cyclic (and possibly fully connected)
potentials in 1-dimensional $m$-partite matching.  The star potential
based on the same distance metric, does not lead to such a simple
solution as the locations of all three points are important and $w(x)$
is no longer piecewise linear.

Now, remaining in 1-dimension, consider the triangle potential 
built from  a quadratic pairwise
potential. The location of all three points are needed to
determine the cost of a match (in contrast to the linear 
distance function), and it is harder to guess a heuristic.
However, numerical experiments hint that the solution $w(x)$ is quadratic
so we try the  anzatz $w(x) = a + b(x-{1\over 2})^2$ (which respects the 
symmetry about the mid-point $x=1/2$) in the maximization equation:
\begin{equation}
w(x) = \max_{x_1,x_2} \left[ 
(x_1-x)^2+(x_2-x)^2+(x_1-x_2)^2
-w(x_1)-w(x_2)\right].
\end{equation}
The maximization equations for right hand side yield
\begin{eqnarray}
(2-b) x_1 -x_2 &=& x -{1\over 2}b \nonumber \\
(2-b) x_2 -x_1 &=& x -{1\over 2}b. 
\end{eqnarray}
A manifestly non-optimal solution to the above is $x_2=x_3$ and to avoid
this  we require the above equation to have more than one solution, this
means that $b=1$ or $b=3$. The choice $b=1$ yields $x=1/2$ which 
is clearly generally not the case. The choice $b=3$ yields
$x+x_1 + x_2 = 3/2$ which can be written as
\begin{equation}
{1\over 3} (x+x_1+x_2) = {1\over 2}.
\end{equation}
meaning that the center of mass of optimal triangles is in the center
of the interval. Using this condition in the maximal equation, all
parameters are determined, we find that $a=0$ and the energy is given
by $3/4$ which can further be checked by numerical iteration of
Eq. (\ref{eqsp}). The condition above is not a full heuristic in that
it does not determine the location of all points in the match when
given just one. However, it is supported by Monte Carlo simulations
where the sum of the coordinates of all matching triangles is a bell
shaped distribution with correct mean, and variance decreasing rapidly
with temperature. We expect this result to generalize to $m$-partite
matching with a cyclic potential with quadratic pairwise $V_P$. It
also generalizes to a star potential with quadratic $V_P$ where we
find energy 1/4.  Moreover, although it is not an exact solution, the
quadratic form of $w(x)$ is a good approximation even a quite elevated
temperatures.

The amenability of quadratic potentials carries over to two
dimensions.  We consider tripartite matching with a triangle potential
in a disc so as to preserve rotational symmetry and take $w({\bf r})$
to be independent of angle.  The function to be maximized is smooth
and by differentiating we find that a symmetric maximum occurs when
the center of mass condition ${\bf r}_1+{\bf r}_2+{\bf r}_3=0$ holds.
As in the 1-dimensional case this is not a full heuristic, however it
is sufficient to determine the energy which is $9/2\pi$. This is
consistent with numerical iteration of Eq. (\ref{eqsp}) at low
temperature, though in $2$ dimensions numerical resolution of the
saddle point at low temperatures is difficult.  Furthermore Monte
Carlo simulations indicate that the center of mass condition indeed
becomes sharper as the temperature is reduced. A specific heuristic
which yields this energy makes greedy matches based on equilateral
triangles with center at the origin.  The same heuristic can be used
in the case of a triangle potential in 2-dimensions based on a linear
distance pairwise potential; it leads to an energy of $2\sqrt{3/\pi}$,
but although this is in good agreement numerical results we have not
been able to prove hat it is the true maximum.

\section{Conclusions}
We have presented a statistical mechanics based approach to
multi-index matching problems with site disorder, specifically when
the points to be matched are randomly distributed in a domain $\cal
D$. The cost functions considered are functions of the relative
distances between the points and are consequently correlated, in
contrast to the random link model. We have analyzed the $m$-partite
version of these problems where each match contains one element from
$m$ distinct sets, each of which can in principle have different probability
distributions in the domain ${\cal D}$.  In the thermodynamic limit
the system is described by $m$ functional order parameters.  We have
concentrated on maximal matching problems as the ground state energy
in this case is extensive with the scaling we employ.

In the special cases where the cost function is sufficiently symmetric
and the probability distributions of each set are the same we find set
symmetric solutions  where the functional order parameter  of each set
is the  same. We have  not found any  evidence for a breaking  of this
symmetry,  an  open  question  is  to  whether  there  are  geometries
(e.g.  effects of  boundaries and  spatial dimensionality)  where this
symmetry is spontaneously broken.

For open chain potentials (which of course includes bipartite matching) 
we find a solution to the saddle point equations 
which gives exactly the same free energy as the traveling salesman problem 
with the same number of links. Furthermore this equivalence holds at
all temperatures and is independent of the precise functional form of the 
pairwise potential from which the chain potential is constructed. 
Interestingly this observation is analogous to an equivalence which 
is conjectured to  hold between Euclidean bipartite matching and the 
Euclidean TSP in the 
zero temperature limit of the minimal problem where the cost function is
scaled to give an extensive ground state energy. 

As in the TSP we find that the zero temperature saddle point equations
can be solved via ansatze inspired by heuristics to find the optimal
match. We have been able to analytically solve the saddle point
equations for the function $w$ in a number of instances of maximal
matching in one and two dimensions.  Also we have numerically verified that the
average optimal match is accurately predicted and that the heuristic
which was used to solve the saddle point equation is indeed that
associated with the optimal match. 
Finally we remark that knowledge of the 
function $w$ does not  completely specify a heuristic however it does 
give some partial, potentially useful, information about the heuristic. 
Indeed we have demonstrated that two completely different heuristics can
give the same average of the optimal match for a simple Euclidean 
tripartite-matching problem in one dimension. 
\bigskip

\noindent{\bf Acknowledgement}: DSD would like to thank the Isaac Newton
Insitute, Cambridge University, where part of this work was carried out 
during the  program  Principles of the  Dynamics of Non-Equilibrium Systems. 
DL acknowledges an EPSRC ``Discipline Hopping Award''.

\pagestyle{plain}

\end{document}